\documentclass[aps,reprint]{revtex4-1}
\usepackage{amsmath}
\usepackage{graphicx}
\usepackage[yyyymmdd,hhmmss]{datetime}
\usepackage{color}

\newcommand{\curlypar}[1]{\left\{ #1 \right\}}

\newcommand{\sfrac}[3][]{\left( \frac{#2}{#3} #1 \right)}

\begin{document}


\title{Dissipative transport in superlattices within the Wigner function formalism}



\author{O. Jonasson}
\email{ojonasson@wisc.edu}
\affiliation{University of Wisconsin-Madison, Madison, Wisconsin 53706, USA}

\author{I. Knezevic}
\email{irena.knezevic@wisc.edu}
\affiliation{University of Wisconsin-Madison, Madison, Wisconsin 53706, USA}

\begin{abstract}
We employ the Wigner function formalism to simulate partially coherent, dissipative electron transport in biased semiconductor superlattices. We introduce a model collision integral with terms that describe energy dissipation, momentum relaxation, and the decay of spatial coherences (localization). Based on a particle-based solution to the Wigner transport equation with the model collision integral, we simulate quantum electronic transport at 10 K in a GaAs/AlGaAs superlattice and accurately reproduce its current density vs field characteristics obtained in experiment.
\end{abstract}
\date{\today}

\maketitle

\section{Introduction}
\label{intro}

The Wigner function (WF) enables the phase-space formulation of quantum mechanics \cite{weyl1927,wigner1932,groenewold1946,moyal1949}. It shares many of the characteristics of the classical phase-space distribution function, but can take negative values in phase-space regions of volume smaller than a few $\hbar$, where the Heisenberg uncertainly relation plays an important role \cite{tatarskii1983,hillery1984,jacoboni2004}. The Wigner function formalism has found use in many fields of physics, such as quantum optics~\cite{leonhardt1997,schleich2002}, nuclear physics~\cite{baker1960,shlomo1981}, particle physics \cite{belitsky2004}, and the semiclassical limits of quantum statistical mechanics~\cite{imre1967}.


The Wigner transport equation (WTE) is the equation of motion for the Wigner function (similar to the semiclassical Boltzmann transport equation) and it has been used to model quantum transport in many semiconductor nanostructures~\cite{frensley1986,kluksdahl1989,jensen1991,biegel1996,Bertoni1999,bordone1999,buot2000,garcia2000,shifren2001,nedjalkov2004,jacoboni2004,querlioz2006,dai2007,barraud2011,wojcik2012,jonasson2014}. However, the exact form of the collision integral for the Wigner transport equation remains an outstanding open question. In the past, the effects of scattering with phonons and impurities in the Wigner transport equation were accounted for via a semiclassical Boltzmann collision operator, which employs transition rates calculated based on Fermi's golden rule ~\cite{querlioz2006}. This approach could be justified for the devices considered (resonant tunneling diodes~\cite{frensley1986,kluksdahl1989,querlioz2006}, double-gate MOSFETs~\cite{querlioz2009}, semiconductor nanowires~\cite{barraud2011}, and carbon nanotube FETs~\cite{querlioz2009}), where throughout most of the device region the electrons could be considered semiclassical, point-like particles with a well-defined kinetic energy vs wave vector relationship $E(k)$ from the envelope function and effective mass approximation. In contrast, electrons throughout a biased superlattice may occupy quasibound states, with no simple relationship between the particle energy and wave vector. Considering that the rates derived based on Fermi's golden rule rely on a well-defined kinetic energy pre- and post-scattering, it is clear that a Boltzmann collision integral fails to capture scattering in superlattices.


In this work, we introduce a model collision integral suitable for analyzing dissipative quantum transport in superlattices within the Wigner function formalism. The collision integral features terms responsible for energy dissipation, an  explicit momentum relaxation term, and a term describing the decay of spatial coherences (localization). We discuss each term, the rationale behind it, its effects on the Wigner function, and the justification for the form. On the example of a GaAs/AlGaAs superlattice at 10 K, we solve the Wigner transport equation using a particle-based method with affinities \cite{shifren2001,querlioz2006}. We show that all three terms are required to accurately reproduce the measured current density vs field characteristics.

\section{The Wigner transport equation for superlattices}

\subsection{The superlattice structure}

\begin{figure}
\centering
\includegraphics[width=3.3in]{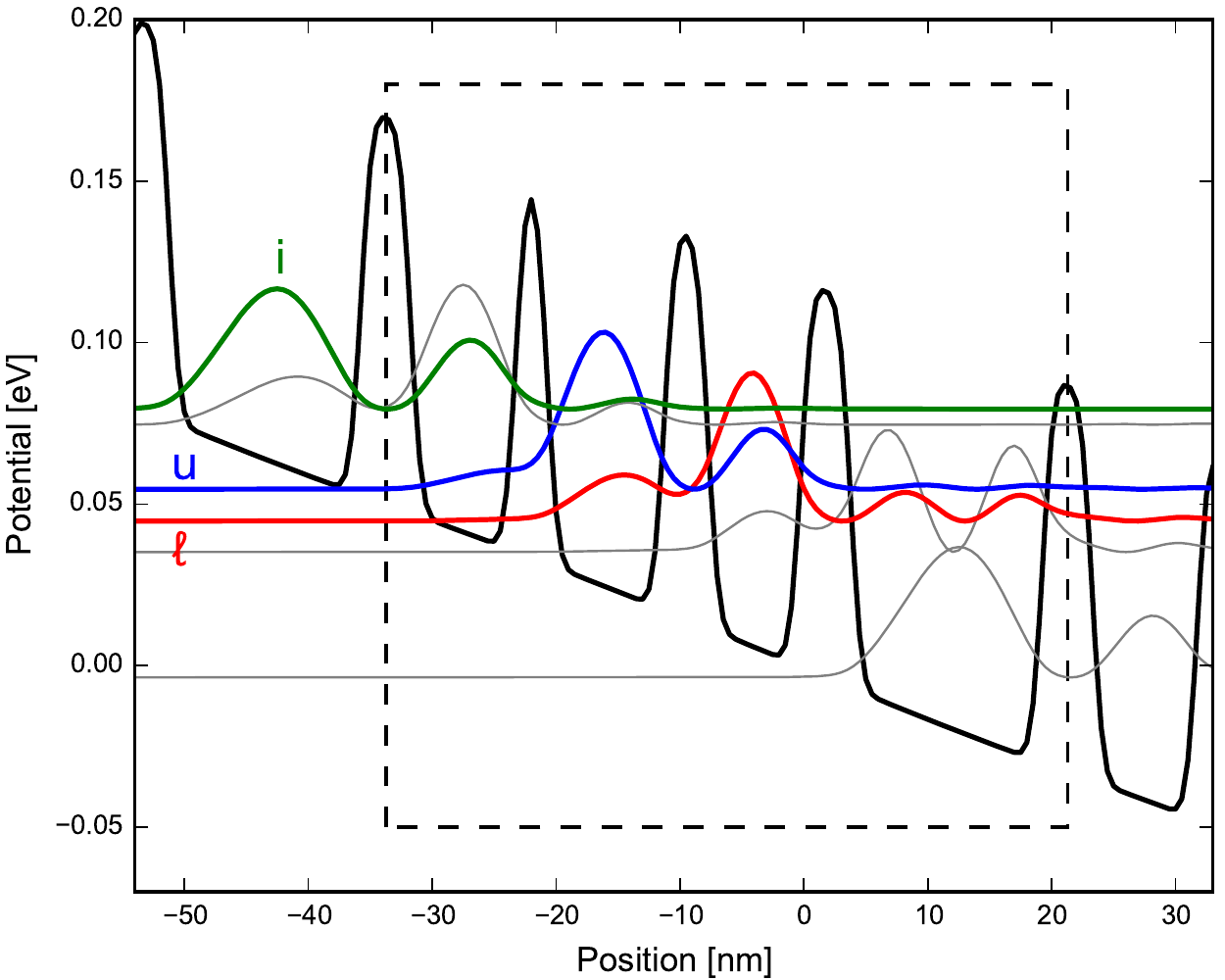}
\caption{Steady-state conduction band profile obtained from the WTE (solid thick black line). The dashed rectangle marks the extent of a single period, where the first barrier is referred to as the injection barrier. Also shown are probability densities corresponding to wavefunction of the upper (u) and lower ($\ell$) lasing levels (thick blue and thick red respectively). Also shown is one of the injector (i) states (thick green) as well as other subbands (thin gray) that are localized near the considered period. Starting at the injection barrier, the layer thicknesses (in nanometers) are
\textbf{4.0}/8.5/\textbf{2.5}/9.5/\textbf{3.5}/7.5/\textbf{4.0}/15.5,
with the widest well n-doped to $2.05\times10^{16}$~cm$^{-3}$ resulting in an average doping of $5.7\times10^{15}$~cm$^{-3}$ per period. (The layer structure has been rounded to the nearest half-nm to match the simulation mesh spacing of $\Delta z=0.5$~nm.) Owing to the low electron concentration, the potential (not including the barriers) is almost linear.}
\label{fig:bandstructure}
\end{figure}

We are interested in simulating electronic transport in biased III-V semiconductor superlattices. Superlattices are systems formed by epitaxial growth of different materials with a certain material sequence reperated tens or hundreds of times, which results in the periodic profile of the conduction and valence band edges in equilibrium, with a period many times greater than the crystalline lattice constant \cite{wacker20021}. Here, we chose an experimentally well-characterized superlattice structure from Ref.~\cite{kumar2011}; this structure has been proposed as a THz quantum cascade laser at a frequency of $1.8$~THz and temperatures up to $168$~K. The structure consists of alternating GaAs (wells) and Al$_{0.15}$Ga$_{0.85}$As (barriers) with a conduction band offset of $0.12$~eV. Figure~\ref{fig:bandstructure} shows the layer structure, as well as the conduction band profile obtained from the WTE simulation coupled with Poisson's equation. The figure also shows several confined states obtained by solving the Schr\"{o}dinger equation at the electric field of $15$~kV/cm (this field is also the design electric field for the laser). The effective mass is $m^*=0.067m_0$, where $m_0$ is the free-electron rest mass.

\subsection{The Wigner transport equation}

In order to simulate time-dependent electronic tranport in a generic semiconductor structure with one-dimensional (1D) transport, we assume translational invariance in the $x-y$ plane perpendicular to the transport direction, $z$. The 1D Wigner function  $f(z,k,t)$ is defined as the Wigner-Weyl transform of the single-particle density matrix $\rho$ via~\cite{wigner1932}
\begin{align}
  \label{eq:wignerdef}
  f(z,k,t)=\frac{1}{2\pi}\int \rho(z+z'/2,z-z'/2) e^{-ikz'} dz'  \ .
\end{align}
The device potential is split into a spatially slowly varying term $V_c(z)$, treated as a classical drift term, and a rapidly varying term $V_q(z)$. In a structure such as a biased superlattice, the applied bias and the Hartree potential are part of the slowly-varying $V_c(z)$, while the barriers are the only terms in the fast-varying $V_q(z)$. This splitting is justified by the low electron concentration in the structure.

The Wigner transport equation describes the time evolution of the WF in the effective mass approximation and can be written as~\cite{jacoboni2001,nedjalkov2004}
\begin{align}
  \label{eq:timeevo}
  \frac{\partial f}{\partial t}=
  -\frac{\hbar k}{m^*}\frac{\partial f}{\partial z}
  +\frac{1}{\hbar}\frac{\partial V_c}{\partial z}\frac{\partial f}{\partial k}
  + Q[V_q,f]+C[f] \ ,
\end{align}
where $C[f]$ is commonly referred to as the collision integral or the collision operator (discussed further below) and $Q[V,f]$ is the quantum evolution term defined by
\begin{align}
  \label{eq:quantumevo}
  Q[V_q,f]=\int_{-\infty}^\infty V_w[V_q](z,k-k')f(z,k') d k' \, .
\end{align}
Here, $V_w[V_q](z,k)$ is the Wigner potential, which can be written as~\cite{nedjalkov2010}
\begin{align}
  \label{eq:wignerpot}
  V_w[V_q](z,k)=\frac{2}{\pi \hbar}\operatorname{Im}\curlypar{ e^{2ikz} \hat V_q(2k) } \ ,
\end{align}
where $\hat V_q(k)$ is the spatial Fourier transform of $V_q$ with the convention
\begin{align}
  \label{eq:hatv}
  \hat V_q(k) = \int_{-\infty}^{\infty} V_q(z) e^{-ikz} d z \ .
\end{align}
%
\subsection{The quantum evolution term for a periodic potential}

As the structure we consider has a periodic fast-varying potential, it can be written as
\begin{align}
  V_q(z)=\sum_{n=-\infty}^{\infty}V_p(z-nL_p) \ ,
\end{align}
where $V_p(z)$ is the potential for a single period and $L_p$ is the period length. The corresponding Wigner potential is given by
\begin{align}
  \label{eq:Vwper}
  V_w[V_q](z,k)=\sum_{n=-\infty}^{\infty}V_w[V_p](z-nL_p,k) \ .
\end{align}
It is tempting to truncate the sum in Eq.~\eqref{eq:Vwper} and only consider a few neighboring periods. However, with this approach, the resulting Wigner potential becomes very ``spiky'', with rapid variation in the $k$-direction and requiring a prohibitively fine mesh. A better approach is to use Eqs.~\eqref{eq:wignerpot} and \eqref{eq:Vwper} and get
\begin{align}
  V_w[V_q](z,k)=\frac{2}{\pi \hbar} \mathrm{Im}
  \left\{ \hat V_q(2k)e^{2izk}\sum_{n=-\infty}^\infty e^{-2inL_pk} \right \} .
\end{align}
Using
\begin{align}
  \sum_{n=-\infty}^\infty e^{-2inL_pk}
  = \frac{\pi}{L_p}\sum_{m=-\infty}^\infty \delta(k-m\pi/L_p)  \ ,
\end{align}
we get
\begin{align}
  V_w[V_q]=\frac{2}{\hbar L_p} \mathrm{Im}
  \left\{\hat V_q(2k)e^{2izk} \right\}\sum_{m=-\infty}^\infty \delta(k-m\pi/L_p) \ .
\end{align}
The quantum evolution term in Eq.~\eqref{eq:quantumevo} then simplifies to
\begin{align}
  \nonumber
  \frac{2}{\hbar L_p} \sum_{m=-\infty}^\infty \mathrm{Im}
    \curlypar{\hat V_q(2m\pi/L_p) e^{2im\pi z/L_p}}f(z,k-m\pi/L_p) \\ \label{eq:Wm}
  = \sum_{m=-\infty}^\infty W_m(z)f(z,k-m\pi/L_p) \ ,
\end{align}
where $W_m$ is a quantum weight, defined as
\begin{align}
  \label{eq:Wm}
  W_m(z)\equiv \frac{2}{\hbar L_p}\mathrm{Im} \curlypar{\hat V(2m\pi/L_p) e^{2im\pi z/L_p}} .
\end{align}
Because the superlattice potential is real, we have $\hat V_q(-k)=\hat V_q^*(k)$. In addition, we have $\hat V(0)=0$, as the bias is included in the slowly varying potential, so the result in Eq.~\eqref{eq:Wm} can be simplified to
\begin{align}
  \nonumber
  Q[V_q,f](z,k,t)=\sum_{m=1}^\infty W_m(z)
  \left[ f(z,k-m\pi/L_p,t) \right.  \\ \label{eq:weightfinal}
  \left. -f(z,k+m\pi/L_p,t) \right] \ .
\end{align}
The definition of the quantum weight $W_m(z)$ is useful because it does not depend on $k$ and can be computed and stored in memory for all relevant values of $m$ and $z$. It only needs to be recomputed when the self-consistent potential is updated. In practice, the infinite sum over $m$ is limited to the values for which $f(z,k\pm m\pi/L_p)$ is non-vanishing.

Here, we will be working with smoothed square barriers, so an analytical expression for the corresponding quantum weight $W_m(z)$ is useful. A single smoothed square barrier of width $2a$ and height $V_0$, centered at the origin, is approximated by (erf is the error function)
\begin{align}
  \label{eq:barrier}
  V_{B}(z) = \frac{V_0}{2} \{ -\operatorname{erf}[\beta(z-a)] + \operatorname{erf}[\beta(z+a)] \}\ .
\end{align}
By performing a Fourier transform we get
\begin{align}
  \hat V_B(k)=\frac{2V_0}{k}e^{-k^2/(4\beta^2)}\sin(ka) \ ,
\end{align}
so
\begin{align}
  W_m[V_B](z)=\frac{2}{\hbar \pi} \frac{V_0}{m}e^{-\frac{m^2\pi^2}{\beta^2 L_p^2}}\sin\sfrac{2\pi ma}{L_p}
  \sin\sfrac{2\pi m z}{L_p} \ .
\end{align}
The above equation represents a single barrier of height $V_0$ centered at the origin. More general structures can be made by using
\begin{align}\label{eq:Wmtot}
  W_m[V](z) =\frac{2}{\hbar \pi}e^{-\frac{m^2\pi^2}{\beta^2 L_p^2}} \sum_{i=1}^{N_B}
 \frac{V_i}{m}\sin\sfrac{2\pi ma_i}{L_p} \\
 \times \sin\sfrac{2\pi m (z-z_i)}{L_p} \ ,\nonumber
\end{align}
where $V_i$, $z_i$ and $2a_i$ are the height, center and width of barrier $i$ and $N_B$ is the total number of barriers in a single period.

Before we continue, we will explain our choice of smoothed barriers. For more smoothing (smaller $\beta$), we can truncate the sum in Eq.~\eqref{eq:Wmtot} at smaller $m$, avoiding the rapidly oscillating high-m terms. This simplification allows use to use a coarser phase-space mesh, speeding up numerical calculations. In this work we used a smoothing length $\beta^{-1}=1.0$~nm. The current-field characteristics are not very sensitive to the value of $\beta$ and we see negligible difference for $\beta^{-1}$ in the range $0.5$ to $1.0$~nm. The choice of smoothed barriers can also be justified by physical reasoning: the lattice constant of GaAs is about $0.5$~nm, which sets the length scale for the most abrupt variation in the conduction band profile.

\subsection{Numerical implementation}

We solve the WTE ~\eqref{eq:timeevo} by using a particle-based method with affinities ~\cite{shifren2001,querlioz2006}.
The WF is written as a linear combination of delta functions on the form
\begin{align}
  \label{eq:faff}
  f(z,k,t)=\sum_p A_p(t)\delta\left[z-z_p(t)\right]\delta\left[k-k_p(t)\right] \ ,
\end{align}
where $A_p(t)$ is the affinity of particle $p$. The diffusive and drift terms [the first and second terms in Eq. \eqref{eq:timeevo}, respectively] represent the free flow and acceleration of particles via a time evolution of the particle position $z_p(t)$ and wave vector $k_p(t)$, respectively. The quantum evolution and collision terms do not affect particle position or wave vector, but induce time evolution of particle affinities given by
\begin{align}
  \label{eq:affevo}
  \sum_{p\in M(z,k)}\frac{dA_p}{dt}=Q[f,V_q](z,k) + C[f] \ ,
\end{align}
where the sum is over all particles $p$ belonging to a mesh point $(z,k)$. We note that by including the collision term in the time evolution of affinities, the particle method is entirely deterministic, and does not require  discretization of the rapidly oscillating diffusive term, which is difficult to deal with using a finite difference method~\cite{kim1999}.

In order to simulate superlattices, periodic boundary conditions are implemented by simply removing particles when they exit the simulation domain and injecting them with the same wave vector at the other side. This approach is valid as long as the simulation domain is large enough that the coherences are not artificially cut off \cite{jonasson2015}. This condition is tested by increasing the number of simulated periods until the simulation converges, typically 100 nm (two periods).

A good choice of the initial state $f(z,k,t=0)$ is important for faster convergence to a steady state, as well as to ensure the steady-state WF is a valid WF ~\cite{tatarskii1983,dias2004}. Here, we will note that, given a valid initial state, the WF will remain valid at later times as long as the model collision integral does not violate the positivity of the density matrix. In this work, we choose the thermal equilibrium WF as the initial state. The steady state is reached in $\sim 1-5$~ps, after which the simulation is continued for additional $\sim 20$~ps to calculate averages of physical observables, such as the current density.

\section{Modeling dissipation and decoherence in superlattices: The collision integral}

In this section, we take a closer look at the collision integral $C[f]$ for the Wigner transport equation, because superlattices pose challenges that are not nearly as dire when addressing other common structures.

Thus far, considerable work has been done on resonant-tunneling diodes (RTDs) using the Wigner transport equation \cite{frensley1986,kluksdahl1989,querlioz2006}. This structure has a relatively small region where tunneling or the formation of bound states is important, and where the relationship between $k$ and the kinetic energy cannot be written. Therefore, throughout most of the simulation domain, the concept of the kinetic energy is well defined and the use of the scattering rates based on Fermi's golden rule is justified.

In contrast, in superlattices, tunneling and quasibound-state formation are important in the entirety of the structure, and the rates calculated based on Fermi's golden rule cannot be reliably used. Employing these rates for superlattices can result in unphysically high currents and negative particle densities. For example, with the use of the common rates for all standard scattering mechanisms \cite{lundstrom2000}, we have been unable to match the experiment for the structure from Fig. 1: the closest match obtained for current densities -- even after allowing for very high electronic temperatures or artificially increased deformation potentials -- is  still several times higher than experiment. Yet, we know that materials parameters are not the issue, because the calculations for the RTDs on the same GaAs/AlGaAs system are quite accurate \cite{querlioz2013}.

In this paper, we abandon the semiclassical rates based on Fermi's golden rule and instead present a simple model collision operator comprising three terms, each with a solid intuitive foundation and no requirement for an $E$ vs $k$ relationship. All three are necessary in order to match the measured current-field characteristics, which is indicative of the interactions that the electrons undergo in superlattices at relatively low temperatures. (We note that a more rigorous form of the collision operator could, for example, be derived based on a Markovian master equation for the single-particle density matrix \cite{knezevicJCEL2013,book:17139} and performing the Wigner-Weyl transform to obtain the WTE; thereby, the Lindblad dissipative term from the master equation would yield the collision operator in WTE.)

The model collision integral we introduce here is given by
\begin{subequations}\label{eq:collision integral}
\begin{eqnarray}
  C[f] =&-& \frac{f(z,k,t)-f_\mathrm{eq}(z,k)}{\tau_R}\label{eq:dissip}\\
&-&\frac{f(z,k,t)-f(z,-k,t)}{\tau_M}\label{eq:mom relax}\\
  &+& \Lambda \frac{\partial^2 f}{\partial k^2} \ . \label{eq:local}
\end{eqnarray}
\end{subequations}
As pointed out earlier, the time evolution induced by the collision integral is included in the time evolution of particle affinities according to Eq.~\eqref{eq:affevo}. The second order $k$-derivative is evaluated using a second-order centered finite-difference scheme.

\subsection{The relaxation term}

The first term, \eqref{eq:dissip}, has the form of the relaxation-time approximation collision integral in semiclassical transport. This term describes relaxation towards equilibrium with a relaxation time $\tau_R$. This term dissipates energy and also partially randomizes momentum.
Even in semiclassical transport, this term technically holds only in the linear regime \cite{lundstrom2000}. It turns out that superlattices have a particular feature that make this approximation -- relaxation to equilibrium as opposed to an \textit{a priori}   unknown steady state --  applicable up to high current densities.

\begin{figure}
\centering
\includegraphics[width=3.3in]{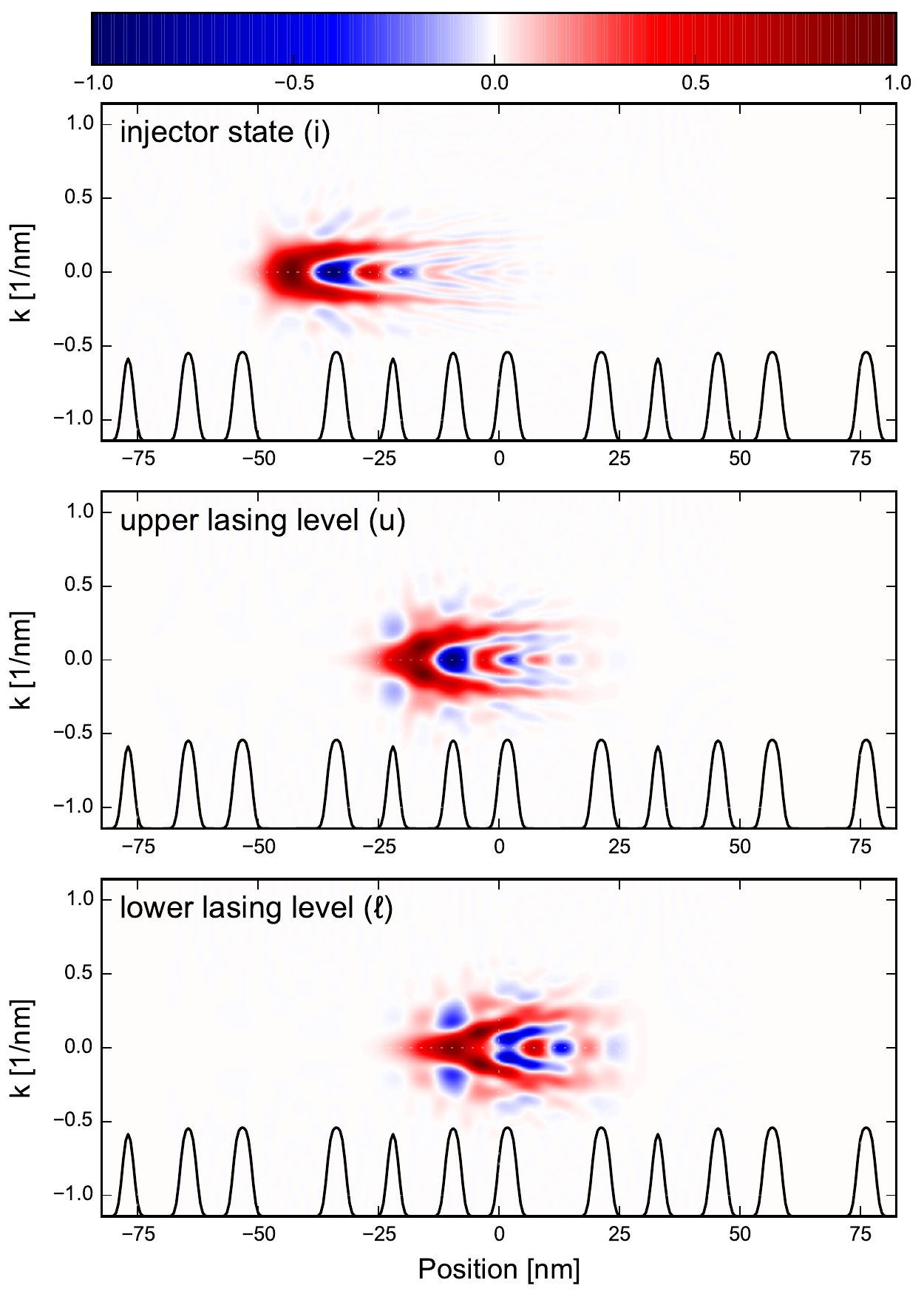}
\caption{Wigner functions for an injector state (i), the upper lasing level (u), and the lower lasing level ($\ell$). See figure \ref{fig:bandstructure} for the spatial profile of the corresponding wave functions.}
\label{fig:popinversion}       
\end{figure}

Figure~\ref{fig:popinversion} shows the WFs for a few relevant subbands (upper lasing, lower lasing, and injector states from Fig.~\ref{fig:bandstructure}) at $10$~K and the electric field of $15$~kV/cm. We find the occupations by taking the subband wavefunctions $\psi_n(z)$ and calculate the corresponding WFs $f_n(z,k)$ by using Eq.~\eqref{eq:wignerdef} with $\rho(z_1,z_2)=\psi_n(z_1)\psi_n^*(z_2)$. The noteworthy feature here is the width of each of the subband WFs in the $k$-direction. The spread over $k$ is of order inverse well width and is considerably greater than the drift wave vector (the approximate shift of the state WFs along the $k$-axis) up to very high current densities. Therefore, for the purpose of forming an approximate collision integral, the nonequilibrium steady-state Wigner function  in superlattice can be approximated by the equilibrium form even under appreciable current.


To calculate the equilibrium Wigner function $f_\mathrm{eq}(z,k)$, we first calculate the equilibrium density matrix. After the Weyl-Wigner transformation in the parallel $\mathbf r_\parallel$ variables and integration over parallel wave vector $\mathbf k_\parallel$, the equilibrium density matrix $\rho_\mathrm{eq}(z_1,z_2)$ is given by
\begin{align}
  \label{eq:rhoeq}
  \rho_\mathrm{eq}(z_1,z_2)=N\sum_s \int_{-\pi/L_p}^{\pi/L_p}\psi_{s,q}(z_1)\psi^*_{s,q}(z_2) \\
  \times \ln (1+e^{ -(E_{s,q}-\mu)/kT } ) d q  \ ,
\end{align}
where $N$ is a normalization factor and $\psi_{s,q}(z)$ are the envelope Bloch wave functions obtained from a solution of the time-independent Schr\"{o}dinger equation for the considered periodic potential. The Fermi level $\mu$ is determined from
\begin{align}
  \label{eq:mu}
  \tilde n = \int D(E)(1+e^{  (E-\mu)/kT })^{-1} d E \ ,
\end{align}
where $\tilde n$ is the electron density averaged over a period and $D(E)$ is the density of states for the superlattice. The equilibrium WF is then calculated using Eq.~\eqref{eq:wignerdef}.
%
%

\subsection{The momentum relaxation term}

In general, momentum and energy relax at different rates, and the discrepancy is accommodated by an explicit term \eqref{eq:mom relax} that randomizes momentum at a rate of $\tau_M^{-1}$, but does not dissipate energy. This term would be very important in the presence of nearly elastic mechanisms, especially if they are efficient at randomizing momentum, such as in the case of acoustic phonons that should be accounted for down to a few Kelvin.  Without this term, current density can be orders of magnitude higher than experimental results, regardless of the choice of the relaxation time $\tau_R$. This term also preserves the positivity of the density matrix.

\subsection{The spatial decoherence (localization) term}

The third term in Eq.~\eqref{eq:collision integral} describes the decay of the spatial coherences of the density matrix, namely,
\begin{equation}
\label{eq:densdeco}
 \Lambda \frac{\partial^2 f(z,k,t)}{\partial k^2} \leftrightarrow -\Lambda (z_1-z_2)^2 \rho(z_1,z_2,t) \ ,
\end{equation}
This term describes the decay of the off-diagonal terms in the density matrix, i.e., spatial decoherence \cite{querlioz2013}, while preserving the positive-definiteness of the density matrix~\cite{diosi1993}. As a result of the action of this term, the density matrix becomes increasingly diagonal over time and we can speak of well-defined position. Therefore, we will use the terms spatial decoherence and localization of the density matrix inerchangaby; $\Lambda$ has previously been referred to as the localization rate \cite{joos2003}, and we adopt the term here as well, even though it is technically a misnomer ($\Lambda$ has the units of inverse length squared and time, rather than the units of inverse time, which a quantity called rate generally has).
In the Wigner function formalism, coherence over large spatial distances corresponds with fast oscillations of the WF in the $k$-direction~\cite{joos2003}. The localization term \eqref{eq:local} depicts diffusion in phase space and tends to smooth out the WF. In simple cases, it is possible to calculate the localization rate $\Lambda$~\cite{dekker1981,joos2003}. Here, we will treat it as a phenomenological parameter that is determined based on comparison with experiment.

There is an additonal, fairly subtle reason why this form is particularly well suited for the description of spatial decoherence in nanostructures that carry current. Namely, nanostructures that carry current are open systems with densely spaced energy levels and generally strong coupling with the contacts \cite{knezevicJCEL2013}. Within the framework of the open systems theory \cite{joos2003,book:17139}, this limit -- smaller energy spacing (per $\hbar$) than the typical relaxation rate of the system -- corresponds to the Brownian motion limit. \footnote{This case is to be contrasted with the optical limit, in which a system is assumed only weakly coupled with the environment and energy levels spaced so far apart that the secular or rotating wave approximation (RWA)  can be emplyed. While tenuous in current-carrying nanostructures, the weak  approximation and RWA are nonetheless often applied to derive master equations in quantum transport \cite{knezevicJCEL2013}.} The spatial decoherence term \eqref{eq:local} has indeed been used to study quantum Brownian motion~\cite{joos2003,book:17139} due to coupling of an open system with an Ohmic bosonic bath by Caldeira and Legget ~\cite{caldeira1983,diosi1993,joos2003}.

\section{Results}
\label{sec:results}

\subsection{Comparison with experiment}

\begin{figure}
\centering
\includegraphics[width=3.3in]{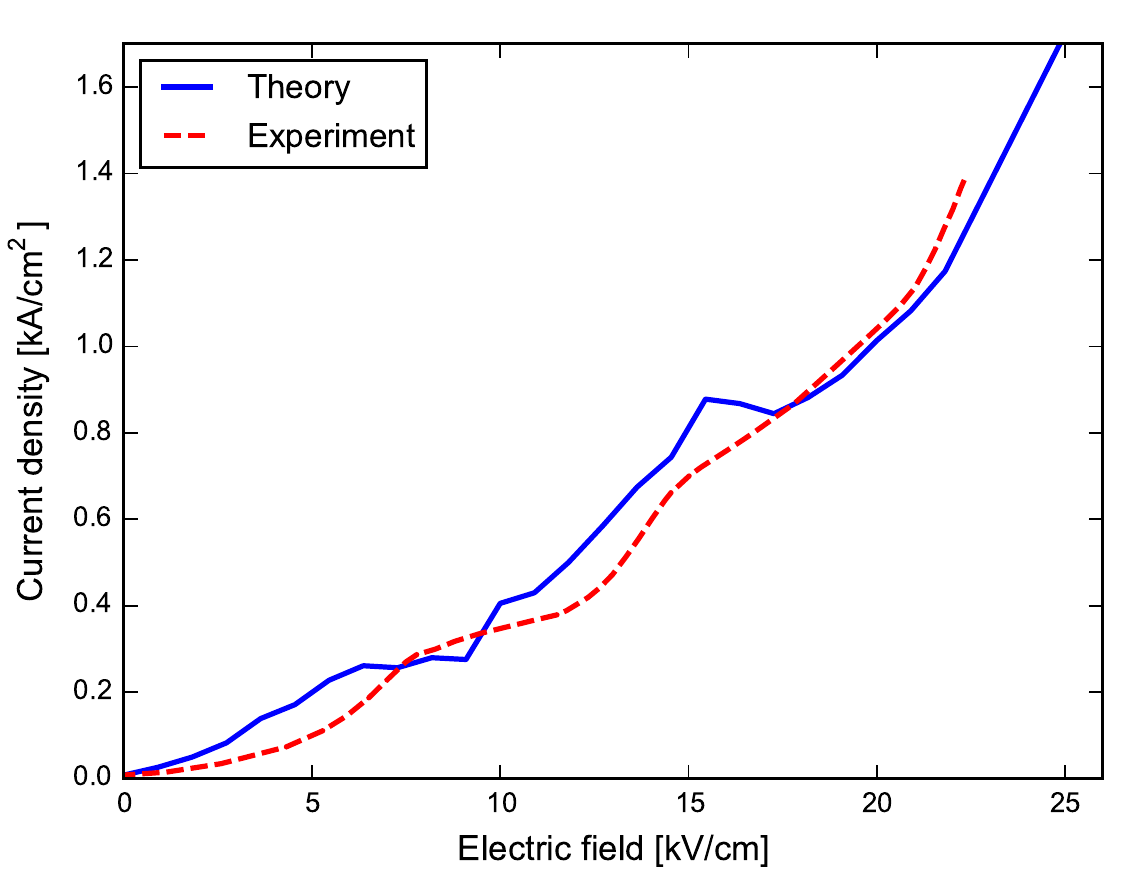}
\caption{Current density vs electric field from theory (solid blue curve) and experimental results from Ref.~\cite{kumar2011} (dashed red curve). $\tau_R^{-1}=10^{12}$~s$^{-1}$, $\tau_M^{-1}=2\times 10^{13}$~s$^{-1}$, and $\Lambda=2\times 10^{9}$~nm$^{-2}$s$^{-1}$.}
\label{fig:JE_comparison}
\end{figure}

Figure~\ref{fig:JE_comparison} shows the calculated current density versus electric field strength and comparison with experiment \cite{kumar2011} for the GaAs/AlGaAs superlattice (Fig. \ref{fig:bandstructure}) at a temperature of $10$~K. The experimental data is for a non-lasing structure (without a waveguide), so electron dynamics is unaffected by the optical field. The experimental $J$--$E$ curve was obtained from the measured $J$--$V$ curve upon incorporating a reported Schottky contact resistance of 4~V \cite{kumar2011}. The best agreement with experiment was achieved with  $\tau_R^{-1}=10^{12}$~s$^{-1}$, $\tau_M^{-1}=2\times 10^{13}$~s$^{-1}$, and $\Lambda=2\times 10^{9}$~nm$^{-2}$s$^{-1}$. The simulation gives current densities very close to experimental results, with plateaus around $8$~kV/cm and $15$~kV/cm.

\begin{figure}
\centering
\includegraphics[width=3.3in]{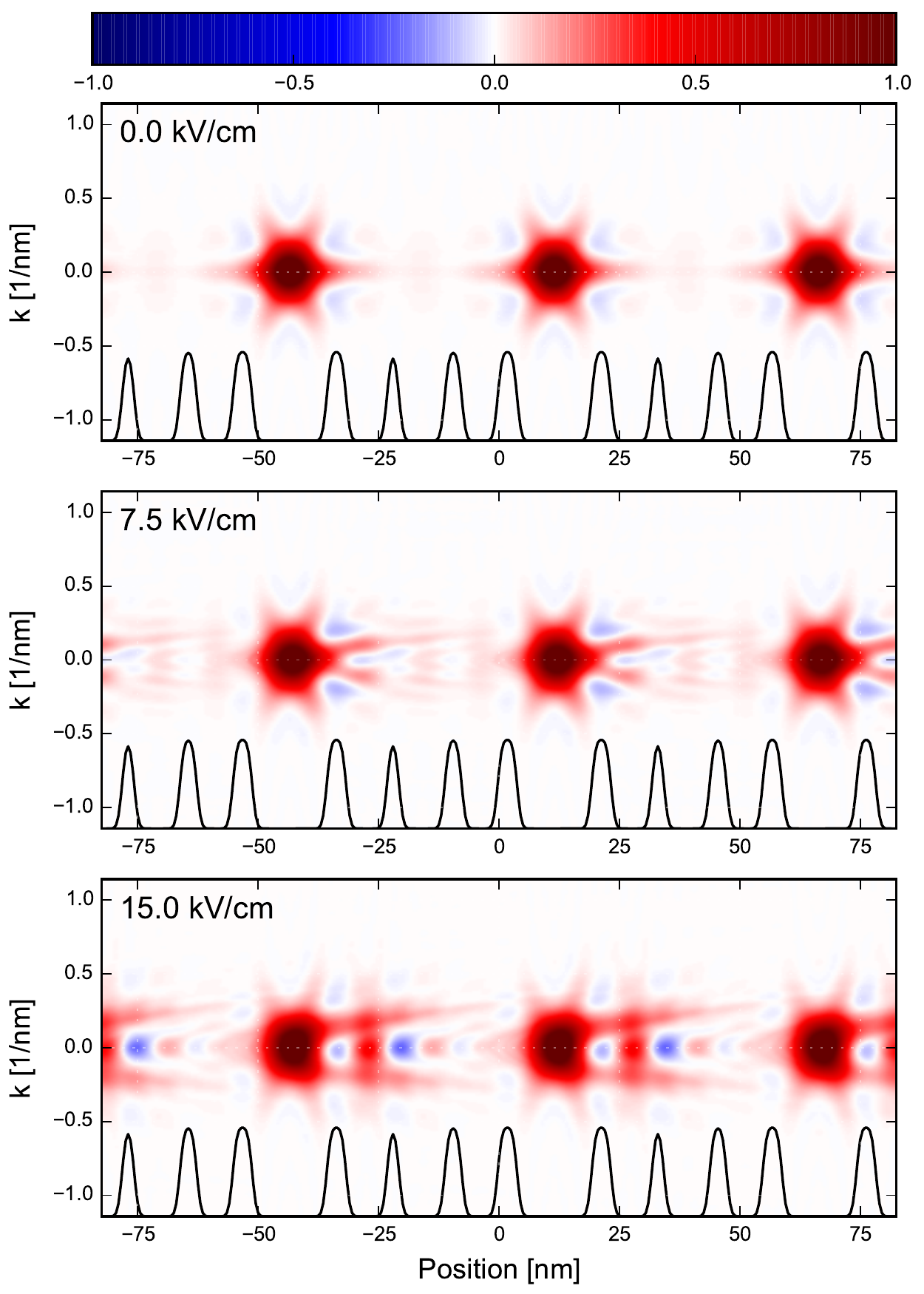}
\caption{Steady state WF for a few values of electric field. The lattice temperature is $10$~K. Barriers are shown in black. All parameters are the same as in Fig~\ref{fig:JE_comparison}.}
\label{fig:triple_wigner}       
\end{figure}

Figure~\ref{fig:triple_wigner} depicts the steady-state WF for different values of the electric field at a lattice temperature of $10$~K, the same temperature and with the same parameters as in Fig. \ref{fig:JE_just_relax}. In the figure we see that, at zero field, due to the low temperature, electrons almost exclusively occupy the ground state in the widest well. When the electric field is increased to $7.5$~kV/cm, tunneling between the widest well and the adjacent well to its right is enhanced. This back-and-fourth motion results in very little net current, as we can see on the corresponding $J$--$E$ curve in Fig. ~\ref{fig:JE_comparison}. At the electric field of $15$~kV/cm, the WF has even higher amplitude in the narrower wells, with pronounced negative values around $k=0$, which is a signature of high occupation of states that are delocalized  between the two wells.

\subsection{Effects of the different terms in the collision integral \eqref{eq:collision integral} on the $J$--$E$ curves}

\begin{figure}
\centering
\includegraphics[width=3.3in]{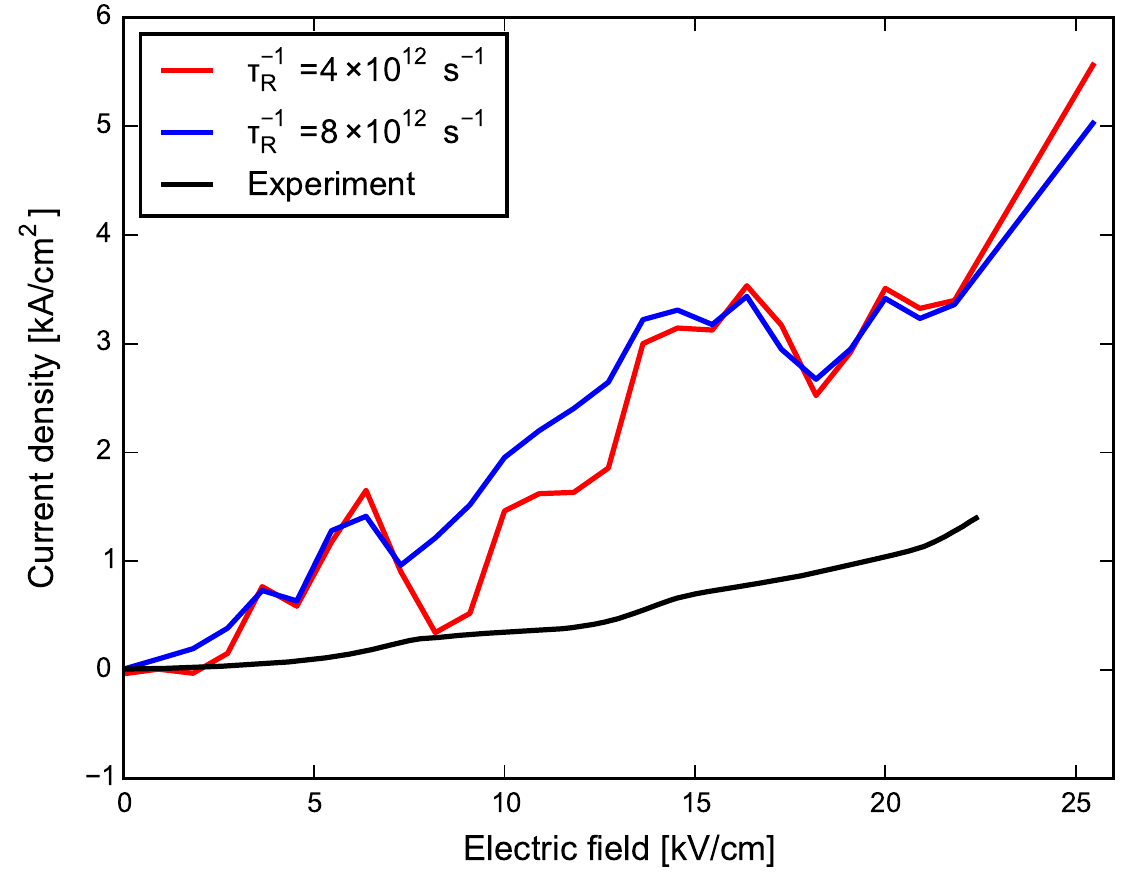}
\caption{Current density vs electric field for different values of the relaxation rate $\tau_R^{-1}$ with $\Lambda=0$ and $\tau_M^{-1}=0$.}
\label{fig:JE_just_relax}   \end{figure}

All three terms in Eq. \eqref{eq:collision integral} are necessary to achieve good agreement with experiment.  In Fig.~\ref{fig:JE_just_relax}, only the relaxation term in \eqref{eq:dissip} is retained and $\tau_R$ is swept ($\tau_M=0$ and $\Lambda=0$). Despite varying $\tau_R$ over many orders of magnitude, we were unable to reproduce experimental results with the term \eqref{eq:dissip} alone for any value of $\tau_R$, with the closest results shown in Fig.~\ref{fig:JE_just_relax}. The current density is overestimated by about an order of magnitude and the $J$--$E$ dependence has features that are absent from the experimental results.

\begin{figure}
\centering
\includegraphics[width=3.3in]{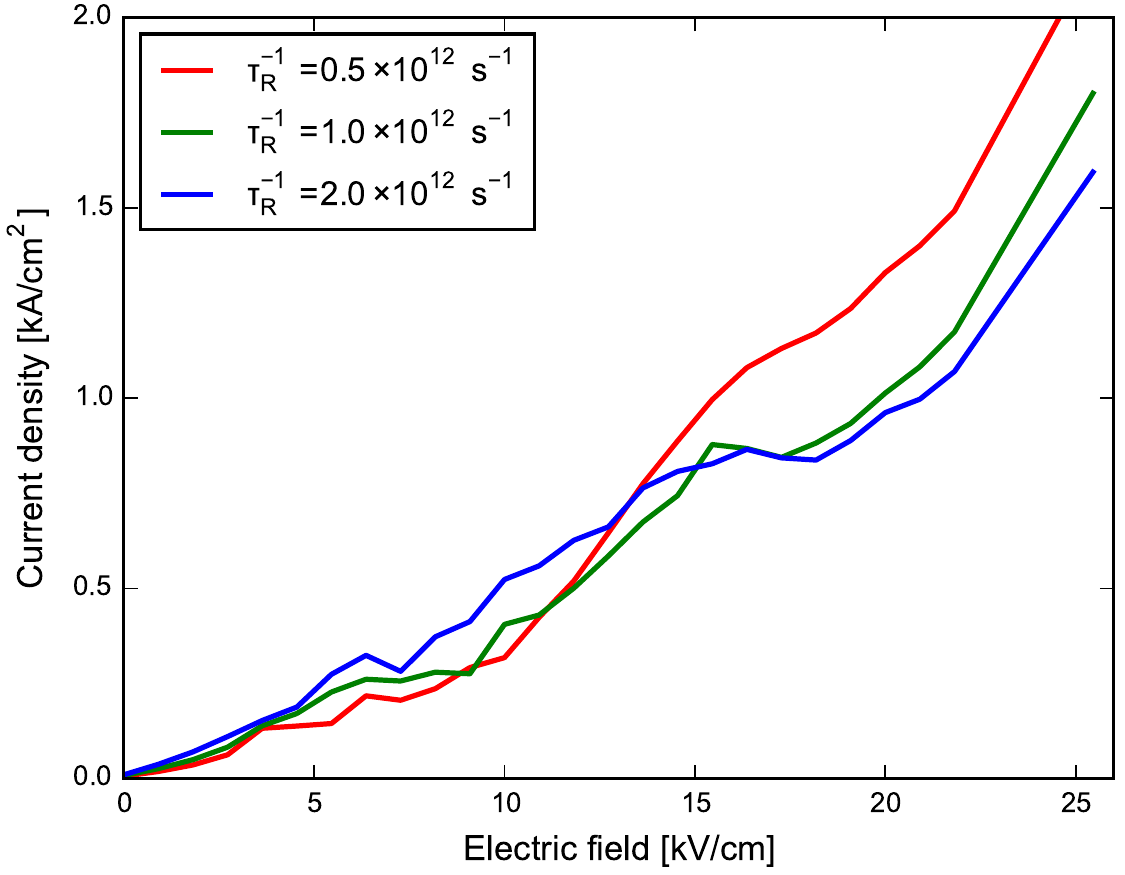}
\caption{Current density vs electric field for different values of the relaxation rate $\tau^{-1}_R$. $\tau_M^{-1}=2\times 10^{13}$~s$^{-1}$ and $\Lambda=2\times 10^{9}$~nm$^{-2}$s$^{-1}$, as in Fig. \ref{fig:JE_comparison}.}
\label{fig:JE_relax}       
\end{figure}

\begin{figure}
\centering
\includegraphics[width=3.3in]{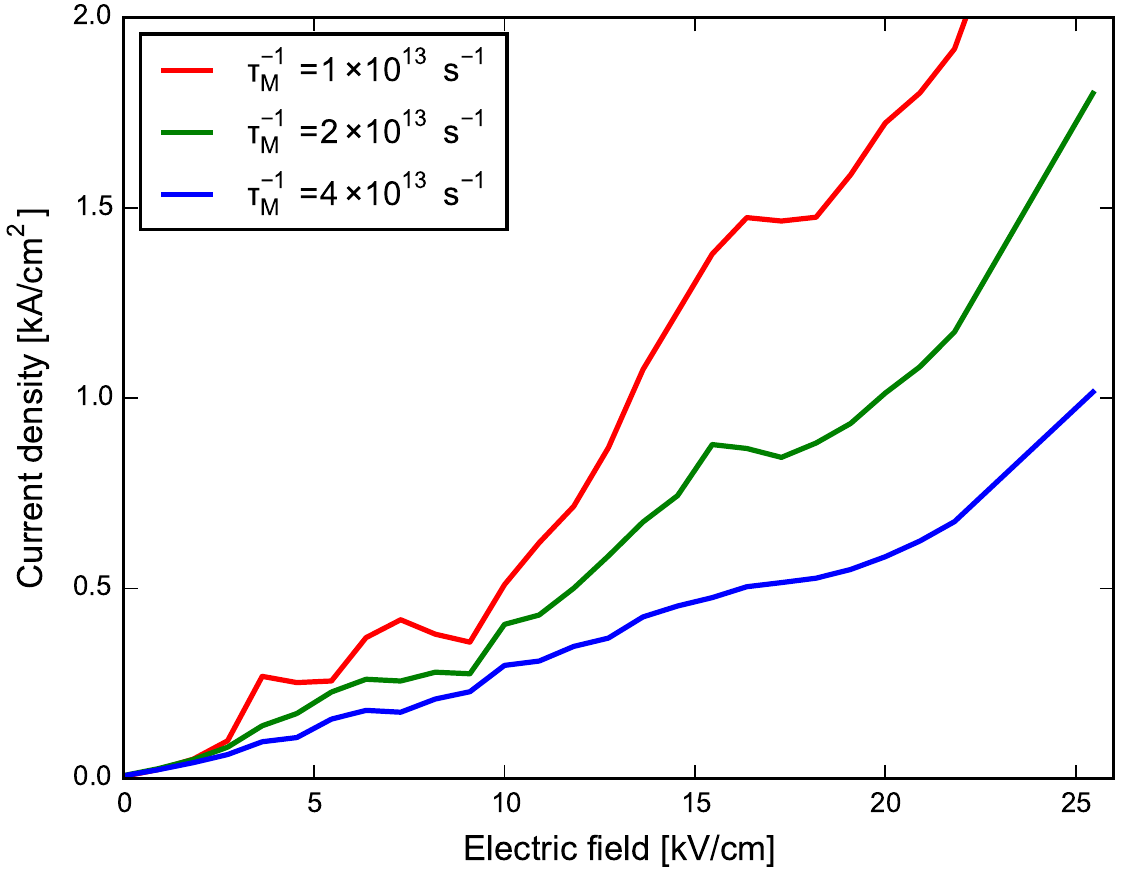}
\caption{Current density vs electric field for different values of the momentum relaxation rate $\tau_M^{-1}$. $\tau_R^{-1}=10^{12}$~s$^{-1}$ and $\Lambda=2\times 10^{9}$~nm$^{-2}$s$^{-1}$, as in  Fig. \ref{fig:JE_comparison}.}
\label{fig:JE_momentum}       
\end{figure}

While the relaxation term \eqref{eq:dissip} is responsible for energy dissipation and for relaxation of the distribution function, it also somewhat relaxes momentum, but not efficiently enough. Figure~\ref{fig:JE_relax} shows the effect of varying the relaxation rate $\tau_R^{-1}$, while Fig.~\ref{fig:JE_momentum} shows the dependence of the current density on the momentum relaxation rate $\tau_M^{-1}$. In each figure, the other two parameters are kept at their best-fit values, as in Fig. \ref{fig:JE_comparison}.  We see that the effect of varying $\tau_R^{-1}$ by a factor of 4 does not have a pronounced effect on the $J$--$E$ curve. However, the $J$--$E$ curve is quite sensitive to the momemtun relaxation rate, with a higher rate resulting in lower current, as expected. A less obvious effect is that a high momentum relaxation rate destroys the fine features in the $J$--$E$ diagram, making the relationship almost linear for $\tau_M^{-1}=4\times 10^{13}$~s$^{-1}$.

\begin{figure}
\centering
\includegraphics[width=3.3in]{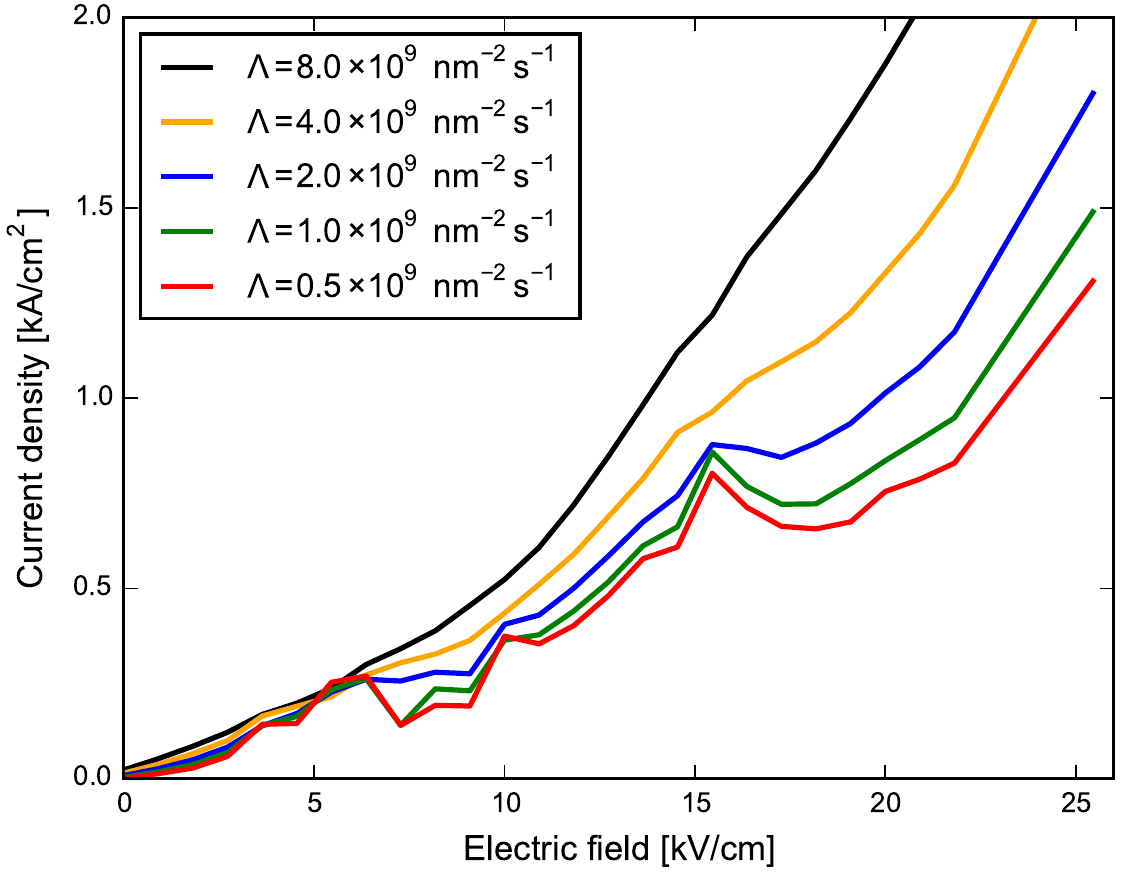}
\caption{Current density vs electric field for different values of the localization rate $\Lambda$, with$\tau_R^{-1}=10^{12}$~s$^{-1}$ and $\tau_M^{-1}=2\times 10^{13}$~s$^{-1}$. High localization rates wash out some of the fine features of the $J$--$E$ curves.}
\label{fig:JE_lambda}       
\end{figure}

Figure~\ref{fig:JE_lambda} shows the dependence of the current density on the localization rate $\Lambda$. As intuitively plausible, lower values of $\Lambda$ result in more prominent fine features in the $J$--$E$ diagram; these get washed out at higher localization rates and the curve is smooth at $\Lambda=8\times 10^{9}$ nm$^{-2}$s$^{-1}$.

\section{Conclusion}

We showed that the Wigner transport equation can be used to model electron transport in semiconductor superlattices. As an example, we considered a GaAs/AlGaAs superlattice of Ref. \cite{kumar2011}, in which transport in partially coherent. The collision integral commonly used with the Boltzmann transport equation is not adequate for use with the WTE in superlattices. Instead, we introduced a model collision integral that comprises three terms: one that captures the dissipation of energy and relaxation towards equilibrium, another that only relaxes momentum, and a third that describes localization, i.e., the decay of spatial coherences due to scattering. The steady-state $J$--$E$ curves are fairly sensitive to the values of the localization and momentum relaxation rates.

Finally, the Wigner equation is exact in the coherent (scattering-free) limit \cite{Sellier20151,Sellier2015254} and its particle-based numerical implementations offer an excellent technique for analyzing ballisitic time-dependent quantum transport in 1D systems (with exciting recent contributions on higher-dimensional systems \cite{SellierJAP13,SellierSISPAD13,SellierCPC14,Ellinghaus2015}). However, the development of systematic approximations for the collision term in the Wigner transport equation is an open problem that presently limits the applications of this intuitive and efficient transport formalism to realistic systems with scattering. This work offers a simple model collision integral that helps accurately capture electronic transport in an experimentally relevant superlattice system, which might enable wider applications of the WTE in transport simulation.

%

\begin{acknowledgements}
The authors gratefully acknowledge support by the U.S. Department of Energy, Office of Basic Energy Sciences, Division of Materials Sciences and Engineering under Award DE-SC0008712. The work was performed using the resources of the UW-Madison Center for High Throughput Computing (CHTC).
\end{acknowledgements}



\end{document}